\documentclass[11pt]{article}

\usepackage[margin=1in]{geometry}
\usepackage{amsmath,amssymb,amsthm}
\usepackage{mathtools}
\usepackage{graphicx}
\usepackage{hyperref}
\usepackage{cite}
\usepackage{enumitem}
\usepackage{authblk}
\usepackage{physics} 
\usepackage{color}
\usepackage{esint}
\hypersetup{
  colorlinks=true,
  linkcolor=blue,
  citecolor=blue,
  urlcolor=blue,
  pdftitle={},
  pdfauthor={}
}

\title{Multi-Matrix Quantum Mechanics, Collective Fields and\\ Emergent Space}

\author[1]{Yue Lei}
\author[2,3]{Suddhasattwa Brahma}
\author[1]{Robert Brandenberger}

\affil[1]{Department of Physics, McGill University, Montr\'eal, QC, H3A 2T8, Canada}

\affil[2]{Physics and Applied Mathematics Unit, Indian Statistical Institute, 203 B.T. Road, Kolkata 700108, India}

\affil[3]{Higgs Centre for Theoretical Physics, School of Physics and Astronomy, University of Edinburgh, Edinburgh, EH9 3FD, UK}

\begin{document}

\date{}
\maketitle

\begin{abstract}
We study quantum mechanics of bosonic multi-matrix Lagragians in the collective field framework, with particular emphasis on three matrix models. We derive the effective Hamiltonian of the collective field and study the vacuum solution and its stability.
\end{abstract}

\tableofcontents


\section{Introduction}

\subsection{Motivation: emergent space from matrix models}

Matrix models have long been proposed as non-perturbative definitions of string theory. In particular, the supersymmetric BFSS matrix model has been proposed \cite{BFSS} as a non-perturbative definition of M-theory, while the IKKT model \cite{IKKT} as a definition of type IIB string theory. These models are mathematically well-defined, consistent quantum mechanical systems whose fundamental degrees of freedom are $N \times N$ Hermitian matrices. At the microscopic level there is no ambient spacetime in the formulation itself, and thus space must arise as an emergent concept. In the IKKT model, which is defined as a matrix integral, even time is not fundamental and must also emerge. In our work, we will focus on quantum mechanical matrix models in which time is present.

This perspective is particularly interesting for cosmological applications. Since matrix models are intrinsically ultraviolet complete quantum systems, they offer the possibility of describing the very early universe without relying on a local low-energy effective field theory in a regime where such a description may break down. This issue is especially important in view of the unitarity \cite{Weiss} and trans-Planckian problmes of effective field theory descriptions of early-universe cosmology \cite{TCC}. In this sense, matrix theory provides a natural framework in which spacetime  \cite{Emergent} and effective field theory can arise only as emergent, approximate notions.

Among supersymmetric matrix models, the BFSS model is distinguished by the existence of a normalizable zero-energy ground state \cite{Graf}. This suggests that an emergent spacetime derived from BFSS may naturally have a vanishing cosmological constant. In addition, there is increasing evidence that although the bosonic sector of BFSS begins with an $SO(9)$ symmetry acting on the nine spatial matrices, the relevant low-energy thermal state may effectively break this symmetry as $SO(9) \rightarrow SO(3)$, so that only three spatial directions become large and classical. Evidence for such a mechanism has been observed in the supersymmetric IKKT model \cite{JN}, its origin has recently been clarified in \cite{Julia}. It is expected that a similar argument exends to the supersymmetric BFSS model as well \cite{Julia2}  (the fact that the $SO(9)$ symmetric state does not minimize the free energy has been established in \cite{Samuel}). Such a symmetry-breaking pattern is precisely what would be required if matrix theory is to describe an emergent $(3+1)$-dimensional universe.

A simple realization of emergent space from matrix dynamics was provided by the single Hermitian matrix model \cite{Sumit}. In that case, the eigenvalues of the matrix provide a one-dimensional emergent space, and in the $N \rightarrow \infty$ limit this space becomes continuous and infinite. The collective field method \cite{Collective} gives a particularly elegant description of this limit, rewriting the dynamics in terms of the eigenvalue density. In the appropriate regime, the resulting collective field theory reproduces the effective spacetime description of the noncritical $c=1$ string \cite{Klebanov}. This remains one of the clearest demonstrations of how spacetime and a continuum field theory can emerge directly from matrix quantum mechanics.

In recent years, several related developments have renewed interest in understanding how geometry is encoded in matrix degrees of freedom. On the one hand, in \cite{us1}, the BFSS model was considered in a high temperature state, and a prescription was given to extract an emergent space from the Matsubara zero modes of the bosonic matrices. The non-zero modes then yield thermal fluctuations which live on the emergent space, and it was shown \cite{us2} that roughly scale-invariant spectra of cosmological fluctuations and gravitational waves emerge, as they do \cite{Nayeri} in the String Gas Cosmology toy model \cite{BV} (see \cite{us3} for a review). On the other hand, there has been significant progress in understanding target-space entanglement and other geometric probes of matrix quantum mechanics \cite{TargetSpaceEntanglement,Hartnoll,CollectiveEntanglementFinite}. These results reinforce the view that geometry is encoded not only in the distribution of matrix eigenvalues but also in the structure of correlations and entanglement among matrix degrees of freedom.

The difficulty, however, is that the single-matrix model is exceptional \cite{Mandal}. For a single Hermitian matrix $X$, one may diagonalize the matrix completely and define the collective density field
\begin{equation}
\varphi(x,t)=\sum_{i=1}^{N}\delta\bigl(x-\lambda_i(t)\bigr),
\label{eq:intro-single-density}
\end{equation}
where the $\lambda_i$ are the eigenvalues of $X$. For multi-matrix models, one may diagonalize only one matrix at a time. The remaining matrices then retain off-diagonal components, and these cannot simply be ignored: from the D-brane point of view, the diagonal entries describe D0-brane positions, whereas the off-diagonal entries represent strings stretched between distinct branes. Conversely, working with the gauge-invariant `loop' variables lead to highly complex Schwinger-Dyson equations, solving which has primarily been dependent on numerical methods \cite{Koch}. A satisfactory, analytical approach to collective-field description of multi-matrix models must therefore be developed to understand emergent space from these models.

In a recent work \cite{Yue}, we took a first step in this direction by studying a two-matrix model. The basic idea was to first fix the gauge by diagonalizing one matrix, then integrate out the off-diagonal modes of the second matrix, and only afterwards pass to collective variables for the remaining diagonal degrees of freedom. This procedure yielded an emergent $(2+1)$-dimensional collective field theory and showed explicitly how the familiar single-matrix construction can be generalized. At the same time, it also revealed an important subtlety: for a BFSS-like model without mass deformation, the induced effective action is generically non-local in time, while a controlled local truncation can be obtained only after introducing a suitable mass scale for the off-diagonal sector \cite{Yue}.

The goal of the present paper is to extend this program to a three-matrix model. This extension is nontrivial. In the two-matrix case, after diagonalizing one matrix, the off-diagonal sector is relatively simple. For three matrices, however, the off-diagonal components of the two non-diagonalized matrices interact with one another, leading to additional mixing terms in the action. The main question is whether these interactions can still be treated in a controlled approximation so that one can derive an effective collective field theory for the diagonal variables alone. A positive answer would be significant, since it would indicate that there is no basic obstruction to extending the construction to larger numbers of matrices, ultimately toward the bosonic sectors of BFSS- and BMN-type models.

In this paper we show that such an extension is indeed possible in a controlled regime. We study a three-matrix model with a BMN-like mass deformation, which provides a natural setting in which the off-diagonal modes are heavy and can be integrated out perturbatively. Working in this framework, we derive the effective collective Hamiltonian for the emergent three-dimensional density field, determine the corresponding large-$N$ vacuum configuration, and analyze the stability of the resulting saddle. In this way, the present work provides the next step in the broader program of deriving emergent spacetime dynamics directly from matrix quantum mechanics.

\subsection{Main results}

The main results of this paper may be summarized as follows:
\begin{itemize}
    \item We formulate a general collective-field construction for bosonic multi-matrix quantum mechanical models, with mass deformations, in a controlled regime.
    \item We study an explicit three-matrix example, with a deformed mass term, and derive the corresponding effective collective Hamiltonian after integrating out the heavy off-diagonal modes.
    \item We obtain the large-$N$ vacuum solution of the resulting three-dimensional collective field theory and analyze quadratic fluctuations about this saddle.
    \item We discuss the extension of the construction to models with $p>3$ matrices, as a step toward a collective-field description of the bosonic sectors of BFSS- and BMN-type matrix models.
\end{itemize}

Beyond these specific results, our broader aim is methodological. We would like to understand whether the combination of gauge fixing, integrating out off-diagonal matrix elements, and only then passing to collective variables can provide a practical route to emergent spacetime in genuinely interacting multi-matrix systems. The present paper gives evidence that this strategy is viable, at least in an approximation regime in which the off-diagonal sector is sufficiently massive.

\subsection{Outline}

The remainder of this paper is organized as follows. In Section~\ref{sec:setup-collective}, we review the collective-field construction for the two-matrix system studied in \cite{Yue}, emphasizing the ingredients that will be needed in the present work. In particular, we recall the gauge-fixing procedure, the diagonal/off-diagonal split of the matrix degrees of freedom, the integration of the off-diagonal modes, and the resulting collective Hamiltonian.

In Section~\ref{sec:three-matrix}, we turn to the three-matrix model that is the main focus of this paper. After fixing the gauge by diagonalizing one of the matrices, we analyze the structure of the remaining diagonal and off-diagonal degrees of freedom and derive the quadratic Hamiltonian governing the off-diagonal sector. We then integrate out these heavy modes and obtain the effective collective Hamiltonian for the emergent three-dimensional density field.

In Section~\ref{sec:vacuum-stability}, we study the large-$N$ vacuum of this collective Hamiltonian. We derive the analytic form of the saddle-point density and then examine quadratic fluctuations around the saddle in order to determine its stability. In Section~\ref{sec:p-greater-than-three}, we discuss the extension of our construction to systems with $p>3$ bosonic matrices and comment on the connection with BFSS/BMN-type models. We conclude in Section~\ref{sec:discussion} with a discussion of the limitations of the present approximation and a brief summary of directions for future work.


\section{Setup: Collective-field review}
\label{sec:setup-collective}

In this section we summarize how the collective-field formalism was developed for a two-matrix model which will serve as the template for the multi-matrix generalization studied below. We keep the discussion self-contained but refer to our previous work \cite{Yue} for detailed derivations.


\subsection{Gauged multi-matrix quantum mechanics}

We consider a $(0+1)$-dimensional $U(N)$ gauged matrix quantum mechanics of $p$ Hermitian matrices $X_I(t)$ in the adjoint representation, coupled to a non-dynamical gauge field $A(t)$ through the covariant derivative, namely
\begin{eqnarray}
 D_t X_I \equiv \dot X_I + [A, X_I] \,, \qquad I = 1,\dots,p \,.
\end{eqnarray}
The action takes the schematic form
\begin{eqnarray}
 S = \frac{1}{2 \ell_s} \int dt \, \mathrm{Tr} \left[ (D_t X_I)^2 - V([X_I]) \right]\,,    
\end{eqnarray}
where $\ell_s$ is a string-length scale and the potential $V([X_I])$ is built from (commutators of) the matrices.  Note that the dimensions of the matrices are chosen to be those of length. The canonical example of this is provided by the BFSS model \cite{BFSS}, given by
\begin{eqnarray}
 S_{\mathrm{BFSS}} = \frac{1}{2 \ell_s} \int dt \, \mathrm{Tr} \left[ (D_t X_I)^2 - \frac{2}{\ell_s^4} [X_I, X_J]^2 + \text{fermions} \right]\,,  
\end{eqnarray}
which includes the supersymmetic partners of the bosonic matrices. However, since our goal is to focus on how spacetime emerges from quantum-mechanical matrix models, we shall restrict our attention to bosonic matrices and their eigenvalue distributions.

Physical states obey the Gauss-law constraint associated with the \(U(N)\) symmetry \cite{Hartnoll},
\begin{eqnarray}
 G \equiv \sum_I [X_I, \Pi_I] \approx 0 \,,
\end{eqnarray}
where the $\Pi_I$ are the canonical momenta of the matrices $X_I$, and we will restrict our attention to the singlet sector. Operationally, this means that the gauge-invariant information about the emergent target space is encoded in traces of products of matrices (Wilson loops), or equivalently, in the eigenvalue distributions of suitably gauge-fixed matrices.

From the D-brane perspective \cite{BFSS}, the diagonal entries of the matrices are interpreted as the positions of $N$ D0-branes in the emergent target space, while the off-diagonal entries represent open strings stretched between these distinct branes. This physical picture in our construction comes from combining both the original collective-field construction \cite{Collective} and more recent analyses of target-space entanglement in gauged multi-matrix models \cite{TargetSpaceEntanglement}.

To illustrate the main features of our construction while still having a tractable system, in our earlier work \cite{Yue} we focused on a two-matrix toy model involving a pair of Hermitian matrices $X$ and $Y$, with the action
\begin{equation}
 S_{2\mathrm{M}} = \frac{1}{2 \ell_s} \int dt \, \mathrm{Tr} \left[ (D_t X)^2 + (D_t Y)^2 - \frac{2}{\ell_s^4} [X,Y]^2 - m^2 Y^2 \right]\,.
 \label{eq:2M-action}
\end{equation}
The quartic commutator reproduces the bosonic part of the Yang-Mills type interaction in the $p=2$ truncation, while the mass term for $Y$ will play a crucial role in rendering the effective description local in time once the ``heavy'' off-diagonal modes are integrated out. However, note that this mass term breaks the symmetry between $X$ and $Y$.


\subsection{Gauge fixing and diagonal/off-diagonal split}

One of the main steps in our algorithm was to first fix the gauge before transforming  to the collective field variables. We work in the so-called ``axial'' gauge:
\begin{eqnarray}
 A(t) = 0\,,
\end{eqnarray}
and use the residual time-independent $U(N)$ symmetry to diagonalize one of the matrices, which we choose to be $X$. Writing
\begin{equation}
 X = \Omega \, \Lambda \, \Omega^{\dagger} \,, \qquad
 \Lambda = \mathrm{diag}(\lambda_1,\dots,\lambda_N) \,, \qquad \Omega \in U(N)\,,
 \label{eq:diag-gauge}
\end{equation}
where we choose the ordering of eigenvalues to be
\begin{eqnarray}
 \lambda_1 < \lambda_2 < \cdots < \lambda_N\,.
\end{eqnarray}
The diagonalization is unique only up to an overall permutation which is fixed by the above ordering convention.

The remaining matrices are rotated into the same basis,
\begin{eqnarray}
 Y = \Omega \, Y^{\mathrm{gf}} \, \Omega^{\dagger} \,,   
\end{eqnarray}
where the superscript ``gf'' denotes the gauge-fixed variables. For the two-matrix model \eqref{eq:2M-action}, we denote
\begin{equation}
 Y^{\mathrm{gf}}_{ij} =
 \begin{cases}
  \rho_i(t) & i=j \, ,\\[2pt]
  Y_{ij}(t) & i \neq j \, ,
 \end{cases}
 \label{eq:Y-split}
\end{equation}
so that $\lambda_i(t)$ and $\rho_i(t)$ label the positions of the $i$-th D0-brane along the two emergent spatial directions, whereas the complex variables $Y_{ij}$, with $i \neq j$ represent the open strings stretching between branes $i$ and $j$.

The change of variables
\begin{eqnarray}
 (X, Y) \longrightarrow (\Lambda, \{Y_{ij}^{\mathrm{gf}}\}, \Omega)
\end{eqnarray}
induces the familiar Vandermonde Jacobian in the path-integral measure. Explicitly, the Haar measure can be written as
\begin{equation}
 dX\, dY = d\Lambda\, dY_{ij}^{\mathrm{gf}}\, d\Omega \, \Delta^2(\lambda) \,,
 \qquad
 \Delta(\lambda) \equiv \prod_{1 \le i < j \le N} (\lambda_i - \lambda_j) \,.
 \label{eq:Vandermonde}
\end{equation}
In the singlet sector the integration over $\Omega$ factorizes, leaving an effective theory for the diagonal eigenvalues and the off-diagonal fluctuations in the gauge-fixed basis.

In the following subsection, we shall integrate out the off-diagonal elements $Y_{ij}$ -- the most crucial step in our program. Following that, we can write down the wavefunction in the diagonal basis \(\Psi(\lambda_i,\rho_i)\). In the Hamiltonian formalism, the Vandermonde factor may be absorbed into this wavefunction. Hence, one can then define 
\begin{equation}
 \Psi(\lambda_i,\rho_i) = \Delta(\lambda) \, \varphi(\lambda_i,\rho_i) \, ,
 \label{eq:Psi-Phi}
\end{equation}
where \(\varphi\) is totally symmetric in the eigenvalues. The transformed Hamiltonian acting on \(\varphi\) is
\begin{equation}
 H_{\mathrm{eff}} = \Delta^{-1}(\lambda) \, H \, \Delta(\lambda) \,,
 \label{eq:Heff-def}
\end{equation}
so that the fermionic statistics of the \(\lambda_i\) (arising from the antisymmetry of \(\Delta\)) is encoded in non-trivial interaction terms in \(H_{\mathrm{eff}}\). This will be the starting point of collective-field theory for multi-matrix quantum mechanics.


\subsection{Integrating out off-diagonal modes}

We now briefly recap how the off-diagonal modes are integrated out in the two-matrix model and a time-local effective action is obtained in a suitable regime.

For clarity, consider first the case $N=2$, where there is a single complex off-diagonal mode $Y_{12}$. After Wick rotation to Euclidean time, $t \to - i \tau$, and expanding the action in components, one finds that the part of the Euclidean action involving \(Y_{12}\) is
\begin{eqnarray}
 S_E[Y_{12}] = \frac{1}{\ell_s} \int d\tau \, \left[ \dot Y_{12}^* \dot Y_{12} + \Omega_{12}(\tau)^2 \, |Y_{12}|^2 \right]\,,
 \qquad
 \Omega_{12}(\tau)^2 \equiv m^2 + \frac{(\lambda_1(\tau) - \lambda_2(\tau))^2}{\ell_s^4}\,.
 \label{eq:Y12-Euclidean}
\end{eqnarray}
Thus, for fixed eigenvalue trajectories $\lambda_i(\tau)$, the off-diagonal field behaves as a harmonic oscillator with a time-dependent frequency set by the brane separation and the mass deformation \cite{TargetSpaceEntanglement}.

Performing the Gaussian functional integral over $Y_{12}$ produces a determinant, which can be written in the standard $`{\rm Tr} \ \log$' form as
\begin{equation}
 Z_{12}[\lambda] \propto \left[ \det \left( \frac{1}{\ell_s} \left\{-\partial_\tau^2 + \Omega_{12}(\tau)^2\right\} \right) \right]^{-1/2} = \exp\left\{ -\frac{1}{2} \, \mathrm{Tr} \log\left(  \frac{1}{\ell_s} \left[-\partial_\tau^2 + \Omega_{12}(\tau)^2 \right]\right) \right\}\,.
 \label{eq:det-Y12}
\end{equation}
This term contributes as an effective potential for the diagonal modes.

If one formally sets $m=0$, the inverse of the kinetic operator generates the Green's function of the massless Euclidean propagator,
\begin{eqnarray}
 \left( -\partial_\tau^2 \right) G_0(\tau,\tau') = \delta(\tau-\tau') \,,
 \qquad
 G_0(\tau,\tau') = - \frac{1}{2} |\tau-\tau'| \,,
\end{eqnarray}
so that expanding the logarithm in \eqref{eq:det-Y12} leads to a non-local-in-time interaction with kernels involving powers of $|\tau-\tau'|$. In this case, there can be no controlled truncation of the resulting infinite series, and the effective action for the diagonal modes is genuinely non-local in time.

By contrast, when $m>0$ the inverse operator has an exponentially decaying Green's function,
\begin{eqnarray}
 \left( -\partial_\tau^2 + m^2 \right) G_m(\tau,\tau') = \delta(\tau-\tau')\,,
 \qquad
 G_m(\tau,\tau') = \frac{1}{2m} e^{- m |\tau-\tau'|}\,,
\end{eqnarray}
so that the expansion of the logarithm in \eqref{eq:det-Y12} can be organized as a derivative expansion in powers of the eigenvalue separation. In particular, in the regime
\begin{equation}
 \frac{1}{m \ell_s^4} \int d\tau \, (\lambda_1(\tau) - \lambda_2(\tau))^2 \ll 1\,,
 \label{eq:locality-regime}
\end{equation}
one may truncate the series at leading order and obtain a time-local effective action whose potential contains a pairwise interaction
\begin{equation}
 V_{\mathrm{eff}}(\lambda_1,\lambda_2) \simeq \frac{1}{4 m \ell_s^4} (\lambda_1 - \lambda_2)^2\,.
 \label{eq:pairwise-potential}
\end{equation}
This is true up to an additive constant that can be interpreted as a cosmological constant term \cite{Yue}, which will be dropped in what follows.

For general $N$, the same analysis applies independently to each off-diagonal component $Y_{ij}$ with $i<j$. In the regime where the analogue of \eqref{eq:locality-regime} holds for all pairs, integrating out all off-diagonal modes yields an effective Euclidean action for the $2N$ diagonal degrees of freedom $(\lambda_i,\rho_i)$, which takes the form
\begin{equation}
 S_{\mathrm{eff}}[\lambda,\rho] \simeq \int d\tau \, \left[ \frac{1}{2 \ell_s} \sum_{i=1}^N (\dot \lambda_i^2 + \dot \rho_i^2 + m^2 \rho_i^2) + \frac{1}{4 m \ell_s^4} \sum_{i<j} (\lambda_i - \lambda_j)^2 \right]\,.
 \label{eq:Seff-diagonal}
\end{equation}
Once again, we have truncated at the leading order term when evaluating the effective action.
Returning to Lorentzian time and passing to the Hamiltonian formulation, one obtains an effective many-body system of $N$ interacting ``fermionic'' eigenvalues $\lambda_i$, interacting through the pairwise potential \eqref{eq:pairwise-potential}, and $N$ bosonic eigenvalues $\rho_i$.


\subsection{Collective variables and Hamiltonian structure}

The collective-field description is obtained by trading the discrete eigenvalues for a continuous density field on the emergent target space. In the two-matrix case, the positions $(\lambda_i,\rho_i)$ of the $N$ D0-branes, corresponding to diagonal elements, define an emergent two-dimensional space with coordinates $(x,y)$. Thus, we introduce the density
\begin{equation}
 \varphi(x,y,t) = \sum_{i=1}^N \delta(x-\lambda_i(t)) \, \delta(y-\rho_i(t))\,,
 \label{eq:density-def}
\end{equation}
subject to the normalization constraint
\begin{equation}
 \int dx \, dy \, \varphi(x,y,t) = N\,.
 \label{eq:density-constraint}
\end{equation}
The field $\varphi(x,y,t)$ thus represents the coarse-grained distribution of D0-branes in the emergent two-dimensional target space.

Upon transforming from the discrete variables $(\lambda_i(t),\rho_i(t))$ to the continuum field $\varphi(x,y,t)$ and its canonically conjugate momentum $\pi(x,y,t)$, the effective Hamiltonian \eqref{eq:Seff-diagonal} can be rewritten in the collective-field form:
\begin{eqnarray}
 H_{\mathrm{coll}} = {} & & \frac{\ell_s}{2} \int dx \, dy \, \Big[ \partial_x \pi(x,y) \, \varphi(x,y) \, \partial_x \pi(x,y)
 + \partial_y \pi(x,y) \, \varphi(x,y) \, \partial_y \pi(x,y) \Big] \nonumber \\[4pt]
 & & + \frac{\ell_s^3 \pi^2}{6} \int dx \, dy \, \varphi(x,y)^3
 + \frac{m^2}{2 \ell_s} \int dx \, dy \, y^2 \varphi(x,y) \nonumber \\[4pt]
 & & + \frac{1}{8 m \ell_s^4} \int dx \, dy \, dx' \, dy' \, \varphi(x,y) \, (x-x')^2 \, \varphi(x',y') 
 - \mu \left( \int dx \, dy \, \varphi(x,y) - N \right)\,,
 \label{eq:H-collective}
\end{eqnarray}
where \(\mu\) is a Lagrange multiplier enforcing the density constraint \eqref{eq:density-constraint}.The first line of \eqref{eq:H-collective} is the kinetic term characteristic of collective-field theories of matrix quantum mechanics \cite{Collective}, written in a manifestly two-dimensional covariant form for the emergent spatial coordinates $(x,y)$. The cubic self-interaction $\sim \varphi^3$ arises from the Vandermonde Jacobian and reduces, upon dimensional reduction, to the standard collective Hamiltonian of the single-matrix model. The mass term $\propto m^2 y^2 \varphi$ originates from the explicit mass deformation in \eqref{eq:2M-action}. Finally, the bi-local term with kernel $(x-x')^2$ is the continuum remnant of the pairwise interaction \eqref{eq:pairwise-potential} generated by integrating out the off-diagonal matrix elements, the strings. This term is local in time but non-local in the emergent spatial direction $x$, encoding a controlled form of spatial non-locality in the collective-field description. Note that this is specific to the Yang-Mills type interaction chosen for our matrix model, and will give rise to an nontrivial structure for the vacuum of the emergent field theory. Since strings are extended objects, it is not surprising that in a point particle field theory truncation non-localities appear.

The two-matrix construction summarized above will be generalized in the rest of the paper to $p>2$ matrices. The key technical ingredients -- gauge fixing to diagonalize one matrix, Gaussian integration of heavy off-diagonal modes in a regime analogous to \eqref{eq:locality-regime}, and rewriting the resulting effective Hamiltonian in terms of a higher-dimensional eigenvalue density $\varphi(\vec{x},t)$ -- are similar to the two-matrix case and provide the starting point for our analysis of vacuum saddles, fluctuations and entanglement in the multi-matrix setup. The main complication, on the other hand, which arises when going from a two matrix system to a $p$ matrix system (with $p \geq 3$) is the interaction between off-diagonal elements of the matrices.

\section{Three-matrix example}
\label{sec:three-matrix}

\subsection{Choice of three-matrix model}

As a minimal setup in which non-commutative vacua and an emergent ``droplet'' description can be studied analytically, we consider a $(0+1)$-dimensional $U(N)$ gauged matrix quantum mechanics with three Hermitian matrices $X^i(t)$ $(i=1,2,3)$ in the adjoint representation. Gauge invariance under
\begin{equation}
X^i \rightarrow U X^i U^\dagger
\end{equation}
is implemented by a non-dynamical gauge field $A_0(t)$ through the covariant derivative
\begin{equation}
D_t X^i \equiv \dot X^i - i [A_0,X^i] \, .
\label{eq:covder3}
\end{equation}

The action is given by
\begin{equation}
S=\frac{1}{2\ell_s}\int dt \, {\rm Tr} \left[ (D_t X^i)^2
-\frac{1}{2\ell_s^4}\sum_i \left( [X^i,X^j]+\ell_s^2  \nu \ \epsilon^{ijk} X^k \right)^2 \right] ,
\label{eq:3M-action}
\end{equation}
where \(\ell_s\) is the string length and \(\nu\) is a mass-deformation parameter. In this
normalization, the matrices \(X^i\) have dimensions of length, while \(\nu\) has dimensions
of mass. This model is a convenient three-matrix analogue of the BMN-type bosonic sector in
the sense that the potential is positive semi-definite and the deformation stabilizes
non-trivial vacua.

The minima of the potential satisfy, up to the standard convention for Hermitian matrix
commutators, the fuzzy-sphere algebra
\begin{equation}
[X^i,X^j]= i \ell_s^2 \nu \epsilon^{ijk} X^k \, ,
\label{eq:fuzzy-sphere}
\end{equation}
which makes the model particularly suitable for studying the emergence of an extended
target-space picture from matrix dynamics. In contrast to the two-matrix system reviewed in
the previous section, the present model already contains the essential new ingredient that
the off-diagonal sectors of the non-diagonalized matrices interact with one another. This is
precisely the feature that makes the three-matrix case the first genuinely non-trivial step
beyond the construction of \cite{Yue}.

Although our long-term motivation is the derivation of an effective collective field theory
for more realistic BFSS/BMN-type systems, it is useful to isolate the minimal case in which
one can still keep analytic control. The model \eqref{eq:3M-action} does exactly this: it is
sufficiently rich to capture mixing between different off-diagonal sectors, while still being
simple enough that the Gaussian integration of heavy modes can be carried out explicitly in
a suitable regime. In what follows, we work in string units and set $\ell_s = 1$.


\subsection{Gauge fixing and diagonal/off-diagonal split}

We now fix the $U(N)$ gauge symmetry by diagonalizing one of the matrices, which we choose
to be $X$. Thus, we write
\begin{equation}
X = {\rm diag}(x_1,\ldots,x_N), \qquad
Y = y + \widetilde Y, \qquad
Z = z + \widetilde Z ,
\label{eq:gauge-fix-3M}
\end{equation}
where
\begin{equation}
y={\rm diag}(y_i), \qquad z={\rm diag}(z_i),
\end{equation}
and the off-diagonal matrices satisfy $(\widetilde Y)_{ii}=(\widetilde Z)_{ii}=0$.  The variables $(x_i,y_i,z_i)$ will eventually play the role of diagonal coordinates in the emergent three-dimensional space, while the off-diagonal elements of $\widetilde Y$ and $\widetilde Z$ correspond to string-like excitations stretched between different branes.

For a diagonal matrix $X$, one has the useful identity
\begin{equation}
[X,M]_{ij}=(x_i-x_j) M_{ij} \equiv \Delta x_{ij} M_{ij}
\qquad (i\neq j) \, ,
\label{eq:XcommM}
\end{equation}
so in particular
\begin{equation}
[X,Y]_{ij}=\Delta x_{ij} y_{ij}, \qquad
[Z,X]_{ij}= - \Delta x_{ij} z_{ij} \, .
\label{eq:basic-comm-3M}
\end{equation}
The only commutator that generates genuine interactions among the off-diagonal modes is
$[Y,Z]$. Decomposing it into pieces linear and quadratic in off-diagonal variables, one finds
for $i\neq j$
\begin{equation}
[Y,Z]_{ij}
=
(y_i-y_j) z_{ij}
-
(z_i-z_j) y_{ij}
+
\sum_{k\neq i,j}
\left(
y_{ik} z_{kj} - z_{ik} y_{kj}
\right) ,
\label{eq:YZ-offdiag}
\end{equation}
while the diagonal part is
\begin{equation}
[Y,Z]_{ii}
=
\sum_{k\neq i}
\left(
y_{ik} z_{ki} - z_{ik} y_{ki}
\right) .
\label{eq:YZ-diag}
\end{equation}

Using \eqref{eq:XcommM}, two of the three perfect-square terms in the potential can be
expanded as
\begin{equation}
{\rm Tr}\left( \nu Z + i[X,Y] \right)^2
=
\nu^2
\left(
\sum_i z_i^2 + 2 \sum_{i<j} |z_{ij}|^2
\right)
+
2 \sum_{i<j} \Delta x_{ij}^{\,2} |y_{ij}|^2
+
i \ 2 \nu \sum_{i\neq j} \Delta x_{ij} y_{ij} z_{ji} ,
\label{eq:pot1-3M}
\end{equation}
and
\begin{equation}
{\rm Tr}\left( \nu Y + i[Z,X] \right)^2
=
\nu^2
\left(
\sum_i y_i^2 + 2 \sum_{i<j} |y_{ij}|^2
\right)
+
2 \sum_{i<j} \Delta x_{ij}^{\,2} |z_{ij}|^2
-
i \ 2 \nu \sum_{i\neq j} \Delta x_{ij} y_{ij} z_{ji} .
\label{eq:pot2-3M}
\end{equation}
The remaining term is most compactly kept in commutator form:
\begin{equation}
{\rm Tr}\left( \nu X + i[Y,Z] \right)^2
=
\nu^2 \sum_i x_i^2
+
2 i \nu \, {\rm Tr}\!\left( X[Y,Z]\right)
-
{\rm Tr}\!\left( [Y,Z]^2 \right) ,
\label{eq:pot3-3M}
\end{equation}
with the components of $[Y,Z]$ given in \eqref{eq:YZ-offdiag} and \eqref{eq:YZ-diag}. When
expanded fully, \eqref{eq:pot3-3M} generates quadratic, cubic and quartic terms in the
off-diagonal variables $(y_{ij},z_{ij})$. Since our immediate aim is to integrate out the
heavy off-diagonal sector, we will isolate the quadratic contribution and treat the higher
terms as interactions. The full polynomial form is recorded in Appendix~\ref{app:three_matrix_potential}. 

At this stage it is worth emphasizing the conceptual difference from the two-matrix case.
In that model, once one diagonalizes one matrix, the off-diagonal modes of the second matrix behave,
to leading order, like independent oscillators with eigenvalue-dependent masses. Here,
however, the off-diagonal modes of $Y$ and $Z$ mix already at quadratic order. This is the
main novelty of the three-matrix problem.


\subsection{Quadratic off-diagonal Hamiltonian}

For the purpose of integrating out heavy off-diagonal modes, we isolate the part of the
Hamiltonian that is quadratic in $(y_{ij},z_{ij})$:
\begin{equation}
H = H_{\rm diag} + H^{(2)}_{\rm off} + H_{\rm int} \, ,
\label{eq:Hsplit3}
\end{equation}
where $H_{\rm diag}$ contains only the diagonal variables, $H^{(2)}_{\rm off}$ contains the
quadratic off-diagonal terms, and $H_{\rm int}$ collects the cubic and quartic contributions.
The kinetic term splits as
\begin{equation}
{\rm Tr}\left( \frac{1}{2}\Pi \cdot \Pi \right)
=
\frac{1}{2}\sum_i
\left(
\pi_{x_i}^2+\pi_{y_i}^2+\pi_{z_i}^2
\right)
+
\sum_{i<j}
\left(
\pi_{y_{ij}}\pi^{*}_{y_{ij}}+\pi_{z_{ij}}\pi^{*}_{z_{ij}}
\right) .
\label{eq:kinetic-split3}
\end{equation}

The quadratic potential for each pair $(i,j)$ can be written compactly as a $2\times 2$
Hermitian mass matrix acting on the vector $(y_{ij},z_{ij})$:
\begin{equation}
V^{(2)}_{\rm off}
=
\sum_{i<j}
\begin{pmatrix}
y_{ij}^{*} & z_{ij}^{*}
\end{pmatrix}
\begin{pmatrix}
2(\nu^2+\Delta x_{ij}^{\,2}+\Delta z_{ij}^{\,2}) &
-\Delta y_{ij}\Delta z_{ij}- i 2 \nu \Delta x_{ij}
\\[4pt]
-\Delta y_{ij}\Delta z_{ij}+ i 2 \nu \Delta x_{ij} &
2(\nu^2+\Delta x_{ij}^{\,2}+\Delta y_{ij}^{\,2})
\end{pmatrix}
\begin{pmatrix}
y_{ij}\\[2pt]
z_{ij}
\end{pmatrix},
\label{eq:mass-matrix-3M}
\end{equation}
where
\begin{equation}
\Delta y_{ij}\equiv y_i-y_j, \qquad
\Delta z_{ij}\equiv z_i-z_j .
\end{equation}
Equation \eqref{eq:mass-matrix-3M} makes clear that the off-diagonal sector is heavy when
$\nu$ is sufficiently large and/or when the diagonal eigenvalues are well separated. In that
regime, the off-diagonal modes can be treated perturbatively and integrated out to leading
order through a Gaussian approximation.

This approximation should be viewed in the same spirit as in the two-matrix case, but with
one important refinement. Because of the mixing between $y_{ij}$ and $z_{ij}$, the heavy
sector is now governed by a matrix-valued kernel rather than a single oscillator frequency.
Nevertheless, as long as the smallest eigenvalue of the mass matrix remains large compared to
the characteristic scales of the slow diagonal dynamics, the same Born-Oppenheimer approximation
continues to apply: the off-diagonal modes respond rapidly and may be integrated out to
obtain an effective description for the ``slow'' diagonal variables alone.


\subsection{Effective Hamiltonian for the collective field}

We now integrate out the off-diagonal modes $(y_{ij},z_{ij})$ to obtain an effective theory
for the diagonal variables $(x_i,y_i,z_i)$, and then rewrite the result in collective-field
form. After Wick rotation, $t\rightarrow -i\tau$, the quadratic off-diagonal sector for each
pair $i<j$ can be written as a Gaussian functional integral over the two-component field
\begin{equation}
\psi_{ij}\equiv
\begin{pmatrix}
y_{ij}\\[2pt]
z_{ij}
\end{pmatrix}
\end{equation}
with kernel
\begin{equation}
S^{(2)}_{\rm off}
=
\sum_{i<j}\int d\tau \,
\psi^{\dagger}_{ij}(\tau)\, A_{ij}(\tau)\, \psi_{ij}(\tau) ,
\qquad
A_{ij}(\tau)=
\left( -\partial_\tau^2 + \nu^2 \right) I_2 + \delta A_{ij}(\tau) ,
\label{eq:Soff-Gaussian}
\end{equation}
where $\delta A_{ij}$ is the pair-dependent operator determined by the quadratic mass matrix
\eqref{eq:mass-matrix-3M}. Performing the Gaussian integral yields, up to an additive
constant,
\begin{equation}
\Delta S_{\rm eff}
=
\frac{1}{2}\sum_{i<j} {\rm Tr}_{\tau}\log A_{ij} .
\label{eq:Trlog3}
\end{equation}

To proceed, we expand around
\begin{equation}
A_0 \equiv \left( -\partial_\tau^2 + \nu^2 \right) I_2 ,
\end{equation}
so that
\begin{equation}
{\rm Tr}_{\tau}\log\left(A_0+\delta A_{ij}\right)
=
{\rm Tr}_{\tau}\log A_0
+
{\rm Tr}_{\tau}\left( A_0^{-1}\delta A_{ij} \right)
+
{\cal O}(\nu^{-2}) .
\label{eq:Trlog-expand3}
\end{equation}
The inverse kernel is
\begin{equation}
A_0^{-1}(\tau-\tau') = G(\tau-\tau') I_2,
\qquad
G(\tau)=\frac{1}{2\nu}e^{-\nu |\tau|} .
\label{eq:Green3}
\end{equation}
Keeping only the leading time-local term gives
\begin{equation}
{\rm Tr}_{\tau}\left( A_0^{-1}\delta A_{ij} \right)
=
\int d\tau \, G(0)\, {\rm tr}_2\!\left( \delta A_{ij}(\tau)\right)
+
{\cal O}(\nu^{-2}\partial_\tau) ,
\qquad
G(0)=\frac{1}{2\nu} .
\label{eq:local-trace3}
\end{equation}
At this order only the $2\times 2$ trace contributes; the $Y$--$Z$ mixing terms have
vanishing ${\rm tr}_2$ and therefore enter only at subleading order in the derivative
expansion. Using \eqref{eq:mass-matrix-3M}, one finds
\begin{equation}
{\rm tr}_2(\delta A_{ij})
=
2(x_i-x_j)^2 + (y_i-y_j)^2 + (z_i-z_j)^2 .
\label{eq:trace-dA}
\end{equation}
Hence the leading induced pairwise interaction is
\begin{equation}
\Delta S^{(1)}_{\rm eff}
=
\int d\tau \sum_{i<j}
\left[
\frac{1}{2\nu}(x_i-x_j)^2
+
\frac{1}{4\nu}(y_i-y_j)^2
+
\frac{1}{4\nu}(z_i-z_j)^2
\right]
+
{\cal O}(\nu^{-2}) .
\label{eq:induced-pairwise3}
\end{equation}

The asymmetry between $x$ and $(y,z)$ is a remnant of the gauge choice in which $X$ is
diagonalized. In another gauge, the corresponding distinguished direction would be permuted.
We therefore interpret this anisotropy as a gauge artefact of the intermediate description,
rather than as a physical breaking of the symmetry of the underlying model\footnote{A fully
gauge-invariant reformulation would require working directly with symmetric
single-trace observables or loop variables. This makes the system analytically untractable, and hence for our purposes, we stick to the gauge-fixed description.}.

Combining \eqref{eq:induced-pairwise3} with the tree-level diagonal part of
\eqref{eq:3M-action}, we obtain the leading effective Euclidean action for the diagonal
variables:
\begin{eqnarray}
S_{\rm eff}[x,y,z]
&\simeq&
\int d\tau
\left[
\frac{1}{2}\sum_{i=1}^{N}
\left(
\dot x_i^2+\dot y_i^2+\dot z_i^2
\right)
+
\frac{\nu^2}{2}\sum_{i=1}^{N}
\left(
x_i^2+y_i^2+z_i^2
\right)\right.\nonumber\\
& & \left. \;\;\;\;\;\;\;\;\;\;
+
\sum_{i<j}
\left(
\frac{1}{2\nu}(x_i-x_j)^2
+
\frac{1}{4\nu}(y_i-y_j)^2
+
\frac{1}{4\nu}(z_i-z_j)^2
\right)
\right] ,
\label{eq:Seff-diag-3M}
\end{eqnarray}
up to an additive constant, corrections of order $\nu^{-2}$ which are generically non-local
in Euclidean time, and further corrections arising from the cubic and quartic off-diagonal
interactions that were neglected in the Gaussian approximation.

We now return to Lorentzian time and pass to the Hamiltonian description. The $3N$
diagonal coordinates $(x_i,y_i,z_i)$ are traded for a collective density in the emergent
three-dimensional target space:
\begin{equation}
\varphi(\mathbf{x},t)
=
\sum_{i=1}^{N}\delta^{(3)}\!\left(\mathbf{x}-\mathbf{x}_i(t)\right),
\qquad
\mathbf{x}_i\equiv (x_i,y_i,z_i) ,
\label{eq:phi3d}
\end{equation}
subject to the normalization constraint
\begin{equation}
\int d^3x \, \varphi(\mathbf{x},t)=N \, .
\label{eq:phi3d-norm}
\end{equation}
As in the two-matrix case, diagonalizing $X$ induces the Vandermonde factor
\begin{equation}
\Delta(x)=\prod_{i<j}(x_i-x_j) ,
\end{equation}
which can be absorbed into the wavefunction and produces the standard collective
self-interaction proportional to $\varphi^3$.

Introducing the conjugate momentum field $\pi(\mathbf{x},t)$, the leading-order collective
Hamiltonian takes the form
\begin{align}
H_{\rm coll}
&=
\frac{1}{2}\int d^3x \, \varphi(\mathbf{x}) \bigl(\nabla \pi(\mathbf{x})\bigr)^2
+
\frac{\pi^2}{6}\int d^3x \, \varphi(\mathbf{x})^3
+
\frac{\nu^2}{2}\int d^3x \, (x^2+y^2+z^2)\varphi(\mathbf{x})
\nonumber\\[4pt]
&\hspace{0.5cm}
+
\frac{1}{4\nu}\int d^3x\, d^3x'\,
\varphi(\mathbf{x})(x-x')^2 \varphi(\mathbf{x}')
+
\frac{1}{8\nu}\int d^3x\, d^3x'\,
\varphi(\mathbf{x})\Bigl[(y-y')^2+(z-z')^2\Bigr]\varphi(\mathbf{x}')
\nonumber\\[4pt]
&\hspace{0.5cm}
-
\mu \left( \int d^3x \, \varphi(\mathbf{x}) - N \right) .
\label{eq:Hcoll3-final}
\end{align}
The first line contains the universal collective kinetic term, the Vandermonde-induced cubic
interaction, and the harmonic confinement inherited from the mass deformation. The second
line is the continuum version of the induced pairwise interaction
\eqref{eq:induced-pairwise3}, generated by integrating out the heavy off-diagonal modes at
leading order in $1/\nu$.

Equation \eqref{eq:Hcoll3-final} will be the starting point for the saddle-point analysis of
the vacuum droplet and its fluctuations carried out in the next section. The main qualitative
novelty compared to the two-matrix model is that, even though the off-diagonal sector was
treated only at Gaussian order, its mixing structure leaves a non-trivial imprint on the
effective collective theory. In particular, the resulting bilocal kernel already encodes the
fact that the emergent geometry is not simply a direct product of three independent
single-matrix directions, but rather arises from a genuinely interacting multi-matrix system. The fact that we had diagonalized $X$ appears in the asymmetry of the emergent $x$-coordinate when compared to $(y,z)$.


\section{Vacuum Solution and its Stability}
\label{sec:vacuum-stability}

\subsection{Analytic vacuum solution}

We now solve for the static large-$N$ vacuum of the three-matrix collective Hamiltonian
derived above, following what was done in the two-matrix model in \cite{Yue}.
Setting the collective momentum to zero, $\pi(\mathbf{x})=0$, the vacuum density $\varphi_0(\mathbf{x})$ minimizes the collective potential subject to the
constraint $\int d^3x\, \varphi_0 = N$.

Writing the density as
\begin{equation}
\varphi = N\rho ,
\qquad
\int d^3x\, \rho = 1 ,
\end{equation}
and rescale coordinates as
\begin{equation}
\mathbf{x}=N^\alpha \tilde{\mathbf{x}},
\qquad
\rho(\mathbf{x})=N^{-3\alpha}\tilde{\rho}(\tilde{\mathbf{x}}).
\end{equation}
The reason for this rescaling is the standard collective-field logic familiar from the approach adopted in \cite{Sumit, Collective}: In the large-$N$ limit the invariant density becomes a classical field, but the support of the distribution generally grows with $N$, and one must therefore determine its $N$-dependence by balancing the competing terms in the collective Hamiltonian. For our model, this means balancing the universal Vandermonde-induced cubic term against the induced bi-local interaction, while the harmonic term and gradient corrections are then tested for sub-leading behavior.

Given this, the dominant potential terms scale as
\begin{equation}
\int d^3x\, \varphi^3 \sim N^{3-6\alpha},
\qquad
\int d^3x\, d^3x'\, \varphi(\mathbf{x})K(\mathbf{x},\mathbf{x}')\varphi(\mathbf{x}')
\sim N^{2+2\alpha},
\qquad
\int d^3x\, r^2 \varphi \sim N^{1+2\alpha},
\label{eq:scalingterms}
\end{equation}
with
\begin{equation}
K(\mathbf{x},\mathbf{x}')
=
(x-x')^2+\frac{1}{2}(y-y')^2+\frac{1}{2}(z-z')^2 .
\label{Kernel}
\end{equation}
Balancing the first two terms fixes
\begin{equation}
3-6\alpha = 2+2\alpha
\qquad \Longrightarrow \qquad
\alpha=\frac{1}{8} .
\end{equation}
Thus the cubic Vandermonde term and the induced bi-local term both scale as
$N^{9/4}$ and control the leading saddle, while the harmonic confinement is subleading ($\sim N^{5/4}$) at large $N$; gradient terms are even smaller still. We therefore solve for the saddle using only these leading contributions and treat the remaining terms as $1/N$ corrections.

Dropping the subleading gradient and harmonic terms, the stationarity condition
\begin{equation}
\frac{\delta H}{\delta \varphi} = 0
\end{equation}
yields the large-$N$ saddle equation
\begin{equation}
\frac{\pi^2}{2}\varphi_0(\mathbf{x})^2
+
\frac{1}{4\nu}\int d^3x'\, 
K(\mathbf{x},\mathbf{x}')\,\varphi_0(\mathbf{x}')
= \mu ,
\label{eq:saddle_p3_leading}
\end{equation}
with $\mu$ the Lagrange multiplier enforcing $\int d^3x \, \varphi_0=N$. This is the $p=3$ analogue of the semicircle saddle in the single-matrix model and matches the form obtained previously in the two-matrix model.

Assume the vacuum is reflection-symmetric under
\begin{equation}
x\leftrightarrow -x,\qquad y\leftrightarrow -y,\qquad z\leftrightarrow -z .
\end{equation}
Then, we can re-write the kernel integrals as
\begin{align}
\int d^3x'\, (x-x')^2\varphi_0(\mathbf{x}')
&=
Nx^2+\int d^3x'\, x'^2\varphi_0(\mathbf{x}')
\equiv Nx^2+N\sigma_x^2, \\
\int d^3x'\, (y-y')^2\varphi_0(\mathbf{x}')
&=
Ny^2+\int d^3x'\, y'^2\varphi_0(\mathbf{x}')
\equiv Ny^2+N\sigma_y^2, \\
\int d^3x'\, (z-z')^2\varphi_0(\mathbf{x}')
&=
Nz^2+\int d^3x'\, z'^2\varphi_0(\mathbf{x}')
\equiv Nz^2+N\sigma_z^2 ,
\end{align}
where the cross terms vanish by symmetry and $\sigma_{x,y,z}^2$ are moments of the
distribution. Plugging this into the saddle equation shows that the non-local integral is
necessarily a quadratic polynomial in $(x,y,z)$, and therefore $\varphi_0^2$ is also a
quadratic polynomial. The solution thus takes the universal square-root form
\begin{equation}
\varphi_0(x,y,z)
=
\sqrt{c_0-c_1\left(x^2+\frac{y^2}{2}+\frac{z^2}{2}\right)},
\qquad
\varphi_0=0 \quad \text{outside the support}.
\label{eq:phi0_sqrt_form}
\end{equation}

It is convenient to parametrize the support by
\begin{equation}
x^2+\frac{1}{2}(y^2+z^2)\le \Lambda^2,
\qquad
\Lambda^2\equiv \frac{c_0}{c_1},
\label{eq:ellipsoid_support}
\end{equation}
so that the solution \eqref{eq:phi0_sqrt_form} becomes
\begin{equation}
\varphi_0(x,y,z)=
\sqrt{c_1}\,
\sqrt{\Lambda^2-x^2-\frac{y^2}{2}-\frac{z^2}{2}},
\qquad
x^2+\frac{1}{2}(y^2+z^2)\le \Lambda^2 .
\end{equation}

The normalization fixes $c_1$ in terms of $\Lambda$. Using the change of variables
\begin{equation}
u=\frac{x}{\Lambda},
\qquad
v=\frac{y}{\sqrt{2}\Lambda},
\qquad
w=\frac{z}{\sqrt{2}\Lambda},
\end{equation}
for which $u^2+v^2+w^2\le 1$ and $dx\,dy\,dz = 2\Lambda^3\,du\,dv\,dw$, we obtain
\begin{equation}
N=\int d^3x\, \varphi_0
=
2\Lambda^4\sqrt{c_1}
\int_{u^2+v^2+w^2\le 1} d^3u\,
\sqrt{1-u^2-v^2-w^2}
=
2\Lambda^4\sqrt{c_1}\,\frac{\pi^2}{4},
\end{equation}
hence
\begin{equation}
\sqrt{c_1}=\frac{2N}{\pi^2\Lambda^4}.
\label{eq:c1_in_terms_Lambda}
\end{equation}

Finally, the coefficient of $x^2$ in the saddle equation \eqref{eq:saddle_p3_leading} fixes $c_1$ in terms of $\nu$
(since the kernel contributes $\frac{N}{4\nu}x^2$ and the cubic term contributes
$\frac{\pi^2}{2}\varphi_0^2$), giving
\begin{equation}
c_1=\frac{N}{2\pi^2\nu}.
\label{eq:c1_in_terms_nu}
\end{equation}
Combining the previous two relations, \eqref{eq:c1_in_terms_Lambda} and \eqref{eq:c1_in_terms_nu}, yields the droplet scale
\begin{equation}
\Lambda=\left(\frac{8\nu N}{\pi^2}\right)^{1/8},
\label{eq:Lambda_scaling}
\end{equation}
and the explicit large-$N$ vacuum density
\begin{equation}
\varphi_0(x,y,z)=
\begin{cases}
\displaystyle
\frac{2N}{\pi^2\Lambda^4}
\sqrt{\Lambda^2-x^2-\frac{y^2}{2}-\frac{z^2}{2}},
&
x^2+\frac{1}{2}(y^2+z^2)\le \Lambda^2, \\[10pt]
0,
&
x^2+\frac{1}{2}(y^2+z^2)>\Lambda^2 .
\end{cases}
\label{eq:phi0_final}
\end{equation}

Equations \eqref{eq:c1_in_terms_Lambda} - \eqref{eq:phi0_final} exhibit the emergent three-dimensional droplet geometry: the support of $\varphi_0$ defines an ellipsoidal region in $(x,y,z)$ whose linear size scales as $N^{1/8}$ in the large-$N$ limit. The ellipsoidal anisotropy between $x$ and $(y,z)$ is inherited from the choice of gauge that diagonalizes $X$ and from the corresponding kernel $K(\mathbf{x},\mathbf{x}')$. Subleading effects in $1/N$, in particular the harmonic term and gradient corrections, will correct the droplet profile and are naturally incorporated in a systematic expansion around this saddle.


\subsection{Fluctuations around the saddle and stability}

We now study small fluctuations about the large-$N$ saddle,
\begin{equation}
\varphi(\mathbf{x},t)=\varphi_0(\mathbf{x})+\delta\varphi(\mathbf{x},t),
\qquad
\pi(\mathbf{x},t)=\delta\pi(\mathbf{x},t),
\end{equation}
where $\varphi_0$ is the static vacuum profile \eqref{eq:phi0_final} and we have used that the ground state has vanishing collective momentum. Expanding the collective Hamiltonian to quadratic order in
fluctuations gives
\begin{equation}
H_{\rm coll}^{(3)}=H[\varphi_0]+H^{(2)}[\delta\varphi,\delta\pi]+{\cal O}(\delta^3),
\end{equation}
with a universal kinetic term
\begin{equation}
H_{\rm kin}^{(2)}
=
\frac{1}{2}\int d^3x\, \varphi_0(\mathbf{x})\,
\bigl(\nabla\delta\pi(\mathbf{x})\bigr)^2 ,
\end{equation}
and a quadratic potential term obtained from the second variation of the static energy
functional.

Keeping only the leading large-$N$ terms, namely the Vandermonde cubic term and the induced
bi-local interaction, the static energy functional is
\begin{equation}
U[\varphi]
=
\frac{\pi^2}{6}\int d^3x\, \varphi(\mathbf{x})^3
+
\frac{1}{8\nu}\int d^3x\, d^3x'\,
\varphi(\mathbf{x})K(\mathbf{x},\mathbf{x}')\varphi(\mathbf{x}')
-
\mu\left(\int d^3x\, \varphi-N\right),
\end{equation}
with
\begin{equation}
K(\mathbf{x},\mathbf{x}')
=
(x-x')^2+\frac{1}{2}\bigl[(y-y')^2+(z-z')^2\bigr].
\end{equation}
Writing $\varphi=\varphi_0+\delta\varphi$, we cab and impose the linearized constraint
\begin{equation}
\int d^3x\, \delta\varphi(\mathbf{x})=0 .
\end{equation}

Expanding the cubic term to second order gives
\begin{equation}
\frac{\pi^2}{6}\int (\varphi_0+\delta\varphi)^3
=
\frac{\pi^2}{6}\int \varphi_0^3
+
\frac{\pi^2}{2}\int \varphi_0^2\,\delta\varphi
+
\frac{\pi^2}{2}\int \varphi_0(\delta\varphi)^2
+
{\cal O}(\delta^3),
\end{equation}
so the local quadratic contribution is
\begin{equation}
\Delta U^{(2)}_{\rm cubic}
=
\frac{\pi^2}{2}\int d^3x\, \varphi_0(\mathbf{x})\,
\bigl(\delta\varphi(\mathbf{x})\bigr)^2
\ge 0,
\end{equation}
since $\varphi_0(\mathbf{x})\ge 0$ on its support. The linear term
$\int \varphi_0^2 \ \delta\varphi$ is cancelled by the linear variation of the bi-local term together with the $\mu$-constraint, because $\varphi_0$ satisfies the saddle equation.

For the bi-local term, the second variation is simply
\begin{equation}
\Delta U^{(2)}_{\rm bi}
=
\frac{1}{8\nu}\int d^3x\, d^3x'\,
\delta\varphi(\mathbf{x})K(\mathbf{x},\mathbf{x}')\delta\varphi(\mathbf{x}') .
\end{equation}
Because $K$ is quadratic in coordinate differences, this quadratic form depends only on a
finite set of low moments of $\delta\varphi$. In particular, using the linearized constraint one
finds
\begin{align}
\int d^3x\, d^3x'\, \delta\varphi(\mathbf{x})(x-x')^2\delta\varphi(\mathbf{x}')
&=
-2\left(\int d^3x\, x\,\delta\varphi(\mathbf{x})\right)^2, \\
\int d^3x\, d^3x'\, \delta\varphi(\mathbf{x})(y-y')^2\delta\varphi(\mathbf{x}')
&=
-2\left(\int d^3x\, y\,\delta\varphi(\mathbf{x})\right)^2, \\
\int d^3x\, d^3x'\, \delta\varphi(\mathbf{x})(z-z')^2\delta\varphi(\mathbf{x}')
&=
-2\left(\int d^3x\, z\,\delta\varphi(\mathbf{x})\right)^2 .
\end{align}
Thus $\Delta U^{(2)}_{\rm bi}$ only acts on the three center-of-mass moments
$\int x\,\delta\varphi$, $\int y\,\delta\varphi$, and $\int z\,\delta\varphi$.

Fixing the droplet to be centered, or equivalently, restricting to fluctuations that do not shift its centroid,
\begin{equation}
\int d^3x\, x\,\delta\varphi
=
\int d^3x\, y\,\delta\varphi
=
\int d^3x\, z\,\delta\varphi
=0,
\end{equation}
implies
\begin{equation}
\Delta U^{(2)}_{\rm bi}=0
\end{equation}
at this leading order. For such fluctuations, the quadratic static energy reduces to the
manifestly non-negative local term above. Including the universal quadratic kinetic term, we
find that in the leading large-$N$ approximation and at fixed center, the fluctuation
Hamiltonian has no tachyonic directions: the droplet solution is a stable saddle.
Subleading corrections perturb the spectrum but do not destabilize the saddle for
sufficiently large $\nu$.

\section{Generalization to $p>3$ matrices}
\label{sec:p-greater-than-three}

Our construction extends straightforwardly to larger numbers of matrices, provided one works
in a regime where the off-diagonal modes are parametrically heavy so that the quadratic
approximation in the off-diagonal sector remains reliable and the resulting functional
determinant admits a controlled time-local expansion. 

The natural long-term target is a
$p=9$ gauged matrix quantum mechanics in which the bosonic part is of the Yang-Mills type with a mass term, plus Myers deformations. A convenient schematic choice is
\begin{equation}
S_{9M}
=
\frac{1}{2\ell_s}\int dt \, {\rm Tr}\left[
(D_t X^I)^2
-\frac{2}{\ell_s^4}[X^I,X^J]^2
-\sum_{I=1}^{9} m_I^2 (X^I)^2
-\frac{i\kappa}{\ell_s^2}\epsilon^{ijk}X^i[X^j,X^k]
\right],
\label{eq:S9M}
\end{equation}
where $i,j,k\in\{1,2,3\}$, $I,J=1,\ldots,9$, and the last term is a Myers coupling
\cite{Myers} which singles out an $SO(3)\subset SO(9)$. For appropriate relations among the
parameters, the $i=1,2,3$ sector can be reorganized into a sum of squares, stabilizing
fuzzy-sphere vacua and explicitly breaking $SO(9)\rightarrow SO(3)\times SO(6)$ \cite{Keshav}. In the supersymmetric BMN model, the coefficients are fixed by supersymmetry; here, by contrast, we view \eqref{eq:S9M} as a bosonic template in which the same basic structural ingredients are present.

The logic of the collective-field construction is then completely analogous to the lower-dimensional cases. One first fixes the gauge by diagonalizing one matrix, say $X^1$. The remaining matrices retain both diagonal and off-diagonal components, and the latter describe strings stretched between the diagonal D0-brane positions. In a regime where the effective masses of these off-diagonal modes are large, one may integrate them out in a Gaussian approximation, thereby obtaining an effective theory for the diagonal variables alone. The main new complication for large $p$ is that the off-diagonal sectors of different matrices mix with each other, as what already happened in the three-matrix example, but there is no obvious conceptual obstruction to carrying out the same procedure systematically.

Gauge-fixing  and then integrating out the heavy modes in
the large-mass regime produces an effective theory for the $9N$ diagonal variables $x^I_{(a)}(t)$, which are naturally interpreted as the positions of $N$ D0-branes in an
emergent nine-dimensional target space. Passing to collective variables,
\begin{equation}
\varphi(\mathbf{x},t)
=
\sum_{a=1}^{N}\delta^{(9)}\!\left(\mathbf{x}-\mathbf{x}^{(a)}(t)\right),
\qquad
\mathbf{x}=(x_1,\ldots,x_9),
\label{eq:phi9}
\end{equation}
with
\begin{equation}
\int d^9x\, \varphi(\mathbf{x},t)=N,
\label{eq:phi9norm}
\end{equation}
one obtains a $(9+1)$-dimensional collective Hamiltonian of the same general form as in the
three-matrix example \eqref{eq:Hcoll3-final}:
\begin{align}
H_{\rm coll}^{(9)} \simeq\;
&\frac12\int d^9x\;\varphi(\mathbf{x})\,(\nabla\pi(\mathbf{x}))^2
+\frac{\pi^2}{6}\int d^9x\;\varphi(\mathbf{x})^3
+\frac12\int d^9x\;\Big(\sum_{I=1}^9 m_I^2 (x^I)^2\Big)\varphi(\mathbf{x})
\nonumber\\
&+\;\frac{1}{2}\int d^9x\,d^9x'\;\varphi(\mathbf{x})\,\mathcal K(\mathbf{x},\mathbf{x}')\,\varphi(\mathbf{x}')
-\mu\!\left(\int d^9x\,\varphi-N\right)
+\cdots,
\label{eq:Hcoll_9_schematic}
\end{align}
Here $\mathcal{K}(\mathbf{x},\mathbf{x}')$ is the kernel induced by integrating out the off-diagonal
modes, while the ellipsis denotes subleading gradient terms and $\mathcal O(\nu^{-2})$ time-nonlocal corrections. As in the three-matrix case, the second term is the universal Vandermonde-induced cubic interaction, and the last two explicit terms encode the one-body confinement and the induced pairwise interactions among the diagonal degrees of freedom.

In the simplest case where the dominant off-diagonal masses are set by a common deformation
scale, the leading kernel $\mathcal{K}$ reduces to a quadratic form in coordinate differences. In this
sense it generalizes both the $(x-x')^2$ kernel of the two-matrix model \cite{Yue} and the anisotropic kernel of the three-matrix model \eqref{Kernel}. As in the lower-dimensional case, the direction singled out by the gauge choice used to diagonalize one matrix can appear with a different coefficient in the effective kernel. This is a gauge artefact of our description Which could be removed by either averaging over gauge choices or by working with fully symmetric, gauge-invariant observables, such as single-trace or loop variables, rather than the eigenvalues of a chosen matrix.

The structure of the large-$N$ vacuum is also expected to generalize in a direct way. If the cubic Vandermonde term and the induced non-local term dominate over gradient corrections, the saddle-point equation again implies a square-root droplet profile of the schematic form
\begin{equation}
\varphi_0(\mathbf{x})
\propto
\sqrt{\Lambda^2-Q(\mathbf{x})},
\label{eq:9droplet}
\end{equation}
where $Q(\mathbf{x})$ is a positive quadratic form determined by the induced kernel and the
mass matrix $m_I^2$. The support $ Q(\mathbf{x})\le \Lambda^2$ then defines a nine-dimensional ellipsoid, providing an explicit realization of an emergent
nine-dimensional target-space region from the diagonal sector of the matrix model. This is
the higher-dimensional analogue of the droplet found in the two- and three-matrix cases.

In a BMN-like setup, the $SO(3)\times SO(6)$ anisotropy can manifest itself both through the
mass parameters $m_I$ and through the Myers-induced structure of the kernel $\mathcal{K}$, suggesting
that three directions behave differently from the remaining six. This is potentially important from the point of view of Matrix Cosmology, since it offers a concrete route toward realizing a distinguished three-dimensional subspace directly in the collective-field description. Of course, to establish such a scenario in a fully controlled setting would require going beyond the schematic treatment given here, and in particular understanding more precisely how the Gaussian approximation to the off-diagonal sector behaves as one approaches the undeformed BFSS limit.

Our main point is, of course, not to say that all technical issues have been solved for $p=9$, but rather that the three-matrix construction appears to capture the essential new ingredient needed for the general multi-matrix problem: the mixing of off-diagonal sectors. Once this is under control, the remaining steps in the derivation of the collective theory follow the same pattern as in the lower-dimensional examples. In this sense, the present work
should be viewed as a concrete step toward a collective-field description of the bosonic sector of BFSS- and BMN-type matrix models.

\section{Discussion and outlook}
\label{sec:discussion}

In this paper we have derived the collective-field theory of a bosonic three-matrix model with a BMN-like mass deformation \cite{BMN, mass-deformed}, working in an approximation in which the off-diagonal matrix elements are treated at Gaussian order. Within this approximation, the resulting collective Hamiltonian provides an effective description of the matrix-model dynamics in terms of the diagonal degrees of freedom, interpreted as an emergent target-space density. The main new feature relative to the two-matrix case is the mixing between off-diagonal sectors of different matrices, and our three-matrix analysis gives a concrete prescription for handling this complication.

The collective-field theory describes the regime in which the diagonal matrix elements are distinct, so that the off-diagonal modes remain sufficiently heavy to be integrated out in a controlled way. In the coincident-point limit, however, this description necessarily breaks down. This is not surprising: the effective collective theory is obtained after truncating the full matrix dynamics, and near coincidence the would-be heavy modes can no longer be treated perturbatively. In this way, we reconcile the fact that an EFT with a time-dependent UV cutoff is not unitary with the fact that the underlying matrix model which remains the fundamental and unitary description.

Thus, the collective-field theory derived here should be viewed as an effective description valid when the Gaussian approximation is reliable, in particular when the mass deformation parameter $\nu$ is sufficiently large. In the case of the BFSS matrix model, it has been shown that the $\nu \to 0$ limit gives a time nonlocal field theory. It would be interesting to go beyond the the Gaussian approximation and study the $\nu \to 0$ limit.

For applications to supersymmetric matrix models, an essential next step is to generalize the
collective-field theory so as to include fermionic degrees of freedom. Since supersymmetry
plays a central role both in BFSS and in BMN, such an extension is required if one wishes to
address the structure of the true ground state and the dynamics of protected sectors in a more
complete way. It would also be interesting to understand whether the inclusion of fermions can
improve the behavior of the effective theory in regimes where the bosonic truncation becomes
subtle.

Another important direction is to study time-dependent and, in particular, cosmological
solutions of the collective-field equations. This could provide a more direct bridge to the
program \cite{us1, us2} in which cosmological space-time emerges from matrix dynamics, and may help clarify
how matter fluctuations and background evolution are encoded in the collective description.
More broadly, such solutions could shed light on the longstanding problem of deriving the
low-energy effective field theory that emerges from matrix-model cosmology \cite{us1, us2}.

The broader motivation for the present work is that matrix models offer a mathematically well-defined framework in which space is not fundamental but emergent. In that context, the
collective-field approach gives a natural language for translating the dynamics of matrix
eigenvalues into an effective field theory on the emergent target space. Our results suggest
that the passage from two to three matrices already captures the essential new ingredients
needed for more realistic multi-matrix systems, and therefore may provide a useful stepping
stone toward a collective description of the bosonic sector of BFSS- and BMN-type models. There remain, of course, several technical issues to be understood better. Among them are a
more systematic treatment of the neglected non-Gaussian corrections from the off-diagonal
sector, a clearer gauge-invariant formulation of the collective variables, and a more detailed
analysis of subleading gradient and non-local terms in the effective Hamiltonian. Progress on
these questions would sharpen the range of validity of the present construction and help
determine to what extent the emergent droplet picture survives beyond leading order.

With these caveats in mind, the present construction provides evidence that
collective-field methods can be extended beyond the single- and two-matrix settings in a
controlled and physically useful way. If so, this opens the possibility of deriving, directly
from matrix quantum mechanics, an effective target-space description that is rich enough to
address questions of emergent geometry, symmetry breaking, and eventually cosmology.


\section*{Acknowledgments}
S.B. is supported in part by a start-up grant from the Indian Statistical Institute. R.B. and Y. L. are suppored in part by an NSERC Discovery Grant and by funds from the Canada Research Chair program. We thank K. Dasgupta and J. Pasiecznik for collaboration on our previous project, and for many discussions. We are also grateful to Sumit Das for answering many questions.

\appendix

\section{Derivation of the three-dimensional collective Hamiltonian}
\label{app:derivation_collective_p3}

In this appendix we sketch the derivation of the collective-field Hamiltonian
used in the main text for the three-matrix model after integrating out the
off-diagonal modes. We work in the gauge in which the matrix \(X\) is
diagonal, so that the remaining diagonal degrees of freedom are
\(\{(x_i,y_i,z_i)\}_{i=1}^N\). The effective many-body Hamiltonian is
\begin{align}
H=\;&
-\frac12\sum_{i=1}^N \frac{1}{\Delta(x)}\frac{\partial^2}{\partial x_i^2}\Delta(x)
-\frac12\sum_{i=1}^N\frac{\partial^2}{\partial y_i^2}
-\frac12\sum_{i=1}^N\frac{\partial^2}{\partial z_i^2}
+\frac{\nu^2}{2}\sum_{i=1}^N (x_i^2+y_i^2+z_i^2)
\nonumber\\[4pt]
&\qquad
+\frac{1}{2\nu}\sum_{i<j}(x_i-x_j)^2
+\frac{1}{4\nu}\sum_{i<j}(y_i-y_j)^2
+\frac{1}{4\nu}\sum_{i<j}(z_i-z_j)^2,
\label{eq:Hmany_appendix}
\end{align}
where
\begin{equation}
\Delta(x)=\prod_{i<j}(x_i-x_j)
\end{equation}
is the Vandermonde determinant. As in the one-matrix model, absorbing
\(\Delta(x)\) into the wavefunction makes the \(x_i\) variables effectively
fermionic, while the \(y_i\) and \(z_i\) directions remain bosonic.

\subsection{Collective variable and canonical momentum}

We introduce the collective density field
\begin{equation}
\varphi(\mathbf{x})
\equiv
\sum_{i=1}^N \delta^{(3)}(\mathbf{x}-\mathbf{x}_i),
\qquad
\mathbf{x}=(x,y,z),
\qquad
\mathbf{x}_i=(x_i,y_i,z_i),
\label{eq:varphi_def_appendix}
\end{equation}
which obeys
\begin{equation}
\int d^3x\,\varphi(\mathbf{x})=N.
\label{eq:varphi_constraint_appendix}
\end{equation}
Its canonically conjugate field is \(\pi(\mathbf{x})\), satisfying
\begin{equation}
[\varphi(\mathbf{x}),\pi(\mathbf{x}')]=i\,\delta^{(3)}(\mathbf{x}-\mathbf{x}').
\end{equation}

The chain rule gives
\begin{equation}
\frac{\partial}{\partial x_i}
=
\int d^3x\;
\frac{\partial \varphi(\mathbf{x})}{\partial x_i}
\frac{\delta}{\delta \varphi(\mathbf{x})}
=
-\int d^3x\;
\partial_x \delta^{(3)}(\mathbf{x}-\mathbf{x}_i)\,
\frac{\delta}{\delta \varphi(\mathbf{x})},
\label{eq:chainrule_x_appendix}
\end{equation}
and similarly
\begin{equation}
\frac{\partial}{\partial y_i}
=
-\int d^3x\;
\partial_y \delta^{(3)}(\mathbf{x}-\mathbf{x}_i)\,
\frac{\delta}{\delta \varphi(\mathbf{x})},
\qquad
\frac{\partial}{\partial z_i}
=
-\int d^3x\;
\partial_z \delta^{(3)}(\mathbf{x}-\mathbf{x}_i)\,
\frac{\delta}{\delta \varphi(\mathbf{x})}.
\label{eq:chainrule_yz_appendix}
\end{equation}

Following the standard collective-field transformation, the Laplacian terms
produce the collective kinetic energy
\begin{equation}
H_{\rm kin}
=
\frac12\int d^3x\;\varphi(\mathbf{x})\,(\nabla\pi(\mathbf{x}))^2
+\cdots,
\label{eq:Hkin_coll_appendix}
\end{equation}
where the ellipsis denotes the extra terms generated by the
Jacobian of the change of variables from particle coordinates to the collective density
field.

\subsection{Derivation of the Vandermonde-induced cubic term}
\label{appsubsec:cubic_term}

We now derive the cubic potential coming from the Vandermonde determinant in
the \(x\)-sector. Writing the many-body wavefunction as
\begin{equation}
\Psi(x_i,y_i,z_i)=\Delta(x)\,\varphi(x_i,y_i,z_i),
\label{eq:psi_delta_phi_appendix}
\end{equation}
the \(x\)-kinetic operator acting on \(\varphi(x_i,y_i,z_i)\) becomes
\begin{equation}
-\frac12\sum_{i=1}^N \frac{1}{\Delta(x)}
\frac{\partial^2}{\partial x_i^2}\Delta(x).
\label{eq:xkin_vand_appendix}
\end{equation}
Using
\begin{equation}
\partial_{x_i}\ln \Delta(x)=\sum_{j\neq i}\frac{1}{x_i-x_j},
\label{eq:dlogdelta_appendix}
\end{equation}
one sees that the transformed Hamiltonian contains singular interactions
generated by the Vandermonde repulsion. In collective-field language, these
terms give rise to
\begin{equation}
H_{\rm Vdm}
=
\frac12\int d^3x\;
\varphi(\mathbf{x})\,u(\mathbf{x})^2
,
\label{eq:Hvdm_nonlocal_appendix}
\end{equation}
where
\begin{equation}
u(\mathbf{x})
\equiv
\fint dx'\;
\frac{\varphi(x',y,z)}{x-x'}.
\label{eq:u_hilbert_appendix}
\end{equation}
Here \(\fint\) denotes the principal value integral.

To reduce the first term to a local cubic potential, it is convenient to
introduce the normalized Hilbert transform
\begin{equation}
(\mathcal H f)(x)
\equiv
\frac{1}{\pi}\,\fint dx'\,\frac{f(x')}{x-x'}.
\label{eq:hilbert_def_appendix}
\end{equation}
Then
\begin{equation}
u(x,y,z)=\pi\,(\mathcal H \varphi)(x,y,z),
\label{eq:u_Hphi_appendix}
\end{equation}
where \(\mathcal H\) acts only on the \(x\)-coordinate, with \(y\) and \(z\)
held fixed.

For each fixed \((y,z)\), one may use the standard one-dimensional identity
\begin{equation}
\int dx\; f(x)\,\big(\mathcal H f(x)\big)^2
=
\frac13\int dx\; f(x)^3,
\label{eq:hilbert_identity_appendix}
\end{equation}
valid for sufficiently regular \(f\). Applying this to
\(f(x)=\varphi(x,y,z)\) slice by slice in the transverse coordinates gives
\begin{align}
\frac12\int d^3x\;\varphi(\mathbf{x})\,u(\mathbf{x})^2
&=
\frac{\pi^2}{2}\int dy\,dz\int dx\;
\varphi(x,y,z)\,\big(\mathcal H\varphi(x,y,z)\big)^2
\nonumber\\[4pt]
&=
\frac{\pi^2}{6}\int dy\,dz\int dx\;\varphi(x,y,z)^3
\nonumber\\[4pt]
&=
\frac{\pi^2}{6}\int d^3x\;\varphi(\mathbf{x})^3.
\label{eq:cubic_from_hilbert_appendix}
\end{align}
Thus the Vandermonde sector gives
\begin{equation}
H_{\rm Vdm}
=
\frac{\pi^2}{6}\int d^3x\;\varphi(\mathbf{x})^3.
\label{eq:Hvdm_local_appendix}
\end{equation}

\subsection{Bosonic gradient terms from the \(y\)- and \(z\)-sectors}
\label{appsubsec:bosonic_gradient_terms}

Besides the Vandermonde-induced cubic term in the \(x\)-sector, the
collective-field transformation of the ordinary bosonic kinetic terms in the
\(y\)- and \(z\)-directions also generates gradient corrections. These arise
from the Jacobian of the change of variables from the particle coordinates
\(\{y_i,z_i\}\) to the density field \(\varphi(\mathbf{x})\). Schematically,
one finds contributions of the form
\begin{equation}
H_{\rm grad}^{(yz)}
\sim
\frac{1}{8}\int d^3x\;
\frac{(\partial_y\varphi)^2+(\partial_z\varphi)^2}{\varphi}.
\label{eq:Hgrad_yz_appendix}
\end{equation}
These are the natural
three-dimensional analogue of the familiar bosonic ``quantum pressure'' term
in collective-field theory.

In the present work we neglect \eqref{eq:Hgrad_yz_appendix} since it is a higher-derivative correction controlling short-distance and edge structure, and we keep only the
leading bulk terms in the large-\(N\) Hamiltonian. 
\subsection{One-body potential}

The quadratic confining term is straightforward to rewrite:
\begin{equation}
\sum_{i=1}^N (x_i^2+y_i^2+z_i^2)
=
\int d^3x\;(x^2+y^2+z^2)\,\varphi(\mathbf{x}),
\label{eq:onebody_rewrite_appendix}
\end{equation}
so that
\begin{equation}
\frac{\nu^2}{2}\sum_{i=1}^N (x_i^2+y_i^2+z_i^2)
=
\frac{\nu^2}{2}\int d^3x\;(x^2+y^2+z^2)\,\varphi(\mathbf{x}).
\label{eq:mass_rewrite_appendix}
\end{equation}

\subsection{Pairwise interaction terms}

The pairwise terms are rewritten using
\begin{equation}
\sum_{i<j} F(\mathbf{x}_i,\mathbf{x}_j)
=
\frac12 \sum_{i,j} F(\mathbf{x}_i,\mathbf{x}_j),
\label{eq:pairwise_identity_appendix}
\end{equation}
whenever \(F(\mathbf{x},\mathbf{x})=0\).
Therefore,
\begin{align}
\frac{1}{2\nu}\sum_{i<j}(x_i-x_j)^2
&=
\frac{1}{4\nu}\int d^3x\,d^3x'\;
\varphi(\mathbf{x})\,(x-x')^2\,\varphi(\mathbf{x}'),
\label{eq:xpair_final_appendix}
\\[4pt]
\frac{1}{4\nu}\sum_{i<j}(y_i-y_j)^2
&=
\frac{1}{8\nu}\int d^3x\,d^3x'\;
\varphi(\mathbf{x})\,(y-y')^2\,\varphi(\mathbf{x}'),
\label{eq:ypair_final_appendix}
\\[4pt]
\frac{1}{4\nu}\sum_{i<j}(z_i-z_j)^2
&=
\frac{1}{8\nu}\int d^3x\,d^3x'\;
\varphi(\mathbf{x})\,(z-z')^2\,\varphi(\mathbf{x}').
\label{eq:zpair_final_appendix}
\end{align}

\subsection{Final form of the Hamiltonian}

Collecting the kinetic term, the truncated Vandermonde contribution
\eqref{eq:Hvdm_local_appendix}, the confining potential, and the bilocal
interaction terms, and imposing the normalization constraint with a Lagrange
multiplier \(\mu\), we arrive at
\begin{align}
H_{\rm coll} =\;
&\frac12\int d^3x\;\varphi(\mathbf{x})\,(\nabla\pi(\mathbf{x}))^2
\;+\;\frac{\pi^2}{6}\int d^3x\;\varphi(\mathbf{x})^3
\;+\;\frac{\nu^2}{2}\int d^3x\;(x^2+y^2+z^2)\,\varphi(\mathbf{x})
\nonumber\\[4pt]
&+\;\frac{1}{4\nu}\int d^3x\,d^3x'\;\varphi(\mathbf{x})\,(x-x')^2\,\varphi(\mathbf{x}')
\;+\;\frac{1}{8\nu}\int d^3x\,d^3x'\;\varphi(\mathbf{x})\Big[(y-y')^2+(z-z')^2\Big]\varphi(\mathbf{x}')
\nonumber\\[4pt]
&-\;\mu\left(\int d^3x\,\varphi(\mathbf{x})-N\right),
\label{eq:Hcoll_final_appendix}
\end{align}
which is the collective Hamiltonian quoted in the main text.

\section{Expansion of the three-matrix potential}
\label{app:three_matrix_potential}

For completeness, we record here the explicit expansion of the bosonic potential
in terms of diagonal and off-diagonal matrix elements.

We decompose the three Hermitian matrices as
\begin{align}
X_{ij} &= x_i\,\delta_{ij}, \\
Y_{ij} &= y_i\,\delta_{ij} + y_{ij}, \\
Z_{ij} &= z_i\,\delta_{ij} + z_{ij},
\end{align}
with $y_{ii}=z_{ii}=0$, and Hermiticity implies
\begin{align}
y_{ji}=y_{ij}^*, \qquad z_{ji}=z_{ij}^*.
\end{align}
We also define
\begin{align}
\Delta x_{ij} \equiv x_i-x_j .
\end{align}

The bosonic potential is
\begin{align}
V=\Tr\!\left[(\nu Z+i[X,Y])^2+(\nu Y+i[Z,X])^2+(\nu X+i[Y,Z])^2\right].
\end{align}
Since $X$ is diagonal, we immediately have
\begin{align}
[X,Y]_{ij}=\Delta x_{ij}\,y_{ij},\qquad
[Z,X]_{ij}=-\Delta x_{ij}\,z_{ij},
\end{align}
while
\begin{align}
[Y,Z]_{ij}
=(y_i-y_j)z_{ij}-(z_i-z_j)y_{ij}
+\sum_{k\neq i,j}\left(y_{ik}z_{kj}-z_{ik}y_{kj}\right).
\end{align}

Using these relations, the first two perfect-square terms become
\begin{align}
\Tr(\nu Z+i[X,Y])^2
&=\nu^2\left(\sum_i z_i^2+2\sum_{i<j}|z_{ij}|^2\right)
+2\sum_{i<j}\Delta x_{ij}^2\,|y_{ij}|^2
+i\,2\nu\sum_{i\neq j}\Delta x_{ij}\,y_{ij}z_{ji},
\\[4pt]
\Tr(\nu Y+i[Z,X])^2
&=\nu^2\left(\sum_i y_i^2+2\sum_{i<j}|y_{ij}|^2\right)
+2\sum_{i<j}\Delta x_{ij}^2\,|z_{ij}|^2
-i\,2\nu\sum_{i\neq j}\Delta x_{ij}\,y_{ij}z_{ji}.
\end{align}
For the remaining term, it is convenient to write
\begin{align}
\Tr(\nu X+i[Y,Z])^2
=\nu^2\sum_i x_i^2
+2i\nu\,\Tr\!\bigl(X[Y,Z]\bigr)
-\Tr\!\bigl([Y,Z]^2\bigr),
\end{align}
whose full expansion is
\begin{align}
\Tr(\nu X+i[Y,Z])^2
={}&\nu^2\sum_i x_i^2
+i\,2\nu\sum_{i\neq j}(x_i-x_j)y_{ij}z_{ji}
\nonumber\\
&+2\sum_{i<j}\Bigl[
(y_i-y_j)^2|z_{ij}|^2+(z_i-z_j)^2|y_{ij}|^2
-(y_i-y_j)(z_i-z_j)\,y_{ij}z_{ji}
\Bigr]
\nonumber\\
&-2\sum_{i\neq j\neq k}(z_i-z_j)\,y_{ji}\bigl(y_{ik}z_{kj}-z_{ik}y_{kj}\bigr)
-2\sum_{i\neq j\neq k}(y_i-y_j)\,z_{ij}\bigl(y_{jk}z_{ki}-z_{jk}y_{ki}\bigr)
\nonumber\\
&+2\sum_{i\neq k,\;j\neq k}|y_{ik}|^2|z_{kj}|^2
-2\sum_{i\neq j}y_{ij}^2z_{ji}^2
\nonumber\\
&-4\sum_{i\neq j\neq k}y_{ij}z_{ji}
\left(y_{ik}z_{kj}-\frac12 z_{ik}y_{kj}\right)
-4\sum_{i\neq j\neq k}z_{ij}y_{ji}
\left(z_{ik}y_{kj}-\frac12 y_{ik}z_{kj}\right).
\end{align}

Combining the three pieces gives the full potential in terms of diagonal and
off-diagonal variables. In particular, unlike the two-matrix model, the
three-matrix model contains not only quadratic but also cubic and quartic
terms in the off-diagonal sector. In the main text, we retain only the
quadratic part when integrating out the heavy off-diagonal modes, and treat
the higher-order terms perturbatively.

\end{document}